 \documentclass[final,5p,times,twocolumn]{elsarticle}

\usepackage{color,array,amsthm}
\usepackage[hyphens]{url}
\usepackage{graphicx}

\usepackage{amssymb}
\usepackage{amsmath}
\usepackage{mathrsfs}
\usepackage{multirow}

\journal{Microprocessors and Microsystems}

\begin{document}

\begin{frontmatter}



\title{A Digital Beamforming Receiver Architecture Implemented on a FPGA for Space Applications} 


\affiliation[labelUAH]{organization={Space Research Group, Department of Automatics, University of Alcalá},
             city={Alcalá de Henares},
             postcode={28805},
             state={Madrid},
             country={Spain}}

\affiliation[labelUPCT]{organization={Universidad Politécnica de Cartagena},
             city={Cartagena},
             postcode={30202},
             state={Murcia},
             country={Spain}}    

\affiliation[labelCRISA]{organization={Airbus CRISA},
        city={Madrid},
        country={Spain}}                

\author[labelUAH]{EDUARDO ORTEGA}


\author[labelUAH]{AGUSTÍN MARTÍNEZ}

\author[labelUPCT]{ANTONIO OLIVA}


\author[labelCRISA]{FERNANDO SANZ}


\author[labelUAH]{OSCAR RODRÍGUEZ}

\author[labelUAH]{MANUEL PRIETO}

\author[labelUAH]{PABLO PARRA}

\author[labelUAH]{ANTONIO DA SILVA}

\author[labelUAH]{SEBASTIÁN SÁNCHEZ}




\begin{abstract}
The burgeoning interest within the space community in digital beamforming is largely attributable to the superior flexibility that satellites with active antenna systems offer for a wide range of applications, notably in communication services. This paper delves into the analysis and practical implementation of a Digital Beamforming and Digital Down Conversion (DDC) chain, leveraging a high-speed Analog-to-Digital Converter (ADC) certified for space applications alongside a high-performance Field-Programmable Gate Array (FPGA). The proposed design strategy focuses on optimizing resource efficiency and minimizing power consumption by strategically sequencing the beamformer processor ahead of the complex down-conversion operation. This innovative approach entails the application of demodulation and low-pass filtering exclusively to the aggregated beam channel, culminating in a marked reduction in the requisite digital signal processing resources relative to traditional, more resource-intensive digital beamforming and DDC architectures. In the experimental validation, an evaluation board integrating a high-speed ADC and a FPGA was utilized. This setup facilitated the empirical validation of the design's efficacy by applying various RF input signals to the digital beamforming receiver system. The ADC employed is capable of high-resolution signal processing, while the FPGA provides the necessary computational flexibility and speed for real-time digital signal processing tasks. The findings underscore the potential of this design to significantly enhance the efficiency and performance of digital beamforming systems in space applications.
\end{abstract}




\end{frontmatter}


\section{INTRODUCTION}

 Digital Beamforming (DBF) represents a pivotal advancement in antenna array technology, facilitating dynamic alteration of the radiation pattern through sophisticated Digital Signal Processing (DSP) techniques. A hallmark of DBF is its capacity to obviate the need for additional analog hardware by directly controlling the phase and amplitude of signals to individual antenna elements via software. This capability not only introduces a paradigm shift towards `smart' antennas but also significantly enhances performance and adaptability through software-driven adjustments, leveraging the intrinsic advantages of beamforming antenna arrays \cite{DBFRestrospective}\cite{PerfEvalDBF}.

Integrating automated algorithms into DBF systems, such as Adaptive Digital Beamforming (ADBF), Space-time Adaptive Processing (STAP), and MIMO operations, presents a formidable approach to mitigating interference and optimizing beam prioritization and throughput for specific applications \cite{wirth2001radar}\cite{1263229STAP}\cite{7060413MIMO}\cite{shrivastava2015combined}. In the realm of DBF, operations including DSP, channelization, frequency translation, modulation, beamforming, sampling, and synthesis are executed within the digital domain, thereby enabling highly efficient processing that can be readily updated or reconfigured in response to evolving mission parameters. The inherent limitations in reconfiguring analog hardware underscore the value of DBF in enhancing the adaptability of satellite communication payloads \cite{sikri2019multi}.

Recent years have witnessed substantial advancements in Analog-to-Digital Converter (ADC) technology, marked by notable improvements in sample rate and power efficiency \cite{4672032ADConverterTrends}\cite{5724625ADSurvey}\cite{1235166ADSpaceDefense}. These developments have been instrumental in evolving space-based array architectures \cite{7389972}, with high-speed ADCs playing a crucial role in digitizing RF signals for subsequent processing by radiation-tolerant digital signal processors equipped with high-throughput arithmetic capabilities, exemplified by the Kintex UltraScale XQRKU060 FPGA \cite{xilinx2020rt}.

Addressing the nuanced requirements of spaceborne phased arrays concerning size, power, and performance, and leveraging contemporary digital technologies, this paper proposes a novel digital signal processing design for a DBF architecture. By strategically positioning the beamformer processing stage prior to the complex down-conversion operation, our design ensures that only the composite channel is subject to demodulation and low-pass filtering. This innovation aims to streamline digital processing resource allocation and power consumption associated with the demodulation and filtering processes across all incoming channels within the antenna array, facilitating a more resource-efficient implementation of DBF architecture capable of accommodating a substantial number of input channels.

Subsequent sections delve into the hardware design and implementation of the DBF receiver and DDC chain, utilizing the EV12AQ60X-ADX-EVM evaluation board furnished by Teledyne e2v Semiconductors \cite{Teledyne}. This platform integrates the EV12AQ600 ADC and the XCKU085 FPGA \cite{KintexUltraScaleFPGAs}, aligning with the radiation-tolerant XQRKU060 in architectural lineage, to manifest a digital signal processing design that markedly economizes on resources relative to conventional DBF architectures \cite{AntRecSysDBF}.

\section{BEAMFORMING STATE OF ART}

Beamforming is a sophisticated process used to extract and interpret information from signals propagating through space via an array of antennas or sensors. This technique is pivotal in discerning the embedded message within signals for communication purposes or determining the Direction Of Arrival (DOA) of signals, a critical parameter in RADAR, SONAR, and similar applications. When applied to Radio Frequency (RF) array antennas, beamforming's primary goal is to enhance the reception or transmission of information from a targeted signal emanating from a specific direction, while simultaneously suppressing interference from non-targeted directions \cite{johnson1992array}.

At its core, beamforming leverages the phenomena of constructive and destructive interference among electromagnetic waves as they traverse through phased array systems. These interference patterns enable the phased adjustment of signal phases across the array elements, tailoring the beam to match the desired radiation pattern. Due to the reciprocity principle, beamforming techniques are applicable for both signal transmission and reception, with this study specifically emphasizing the implementation of a digital beamforming receiver \cite{665}.

To model the signal propagation through a phased antenna array effectively, we consider the source to be point-like relative to the array's distance, ensuring that it lies within the `far field'—a regime where the wavefront can be approximated as a plane wave. This approximation simplifies the analysis of wave propagation and its interaction with the antenna array \cite{PhasedArrayAntennaHandbook}.

\begin{figure}
\centerline{\includegraphics[width=20pc]{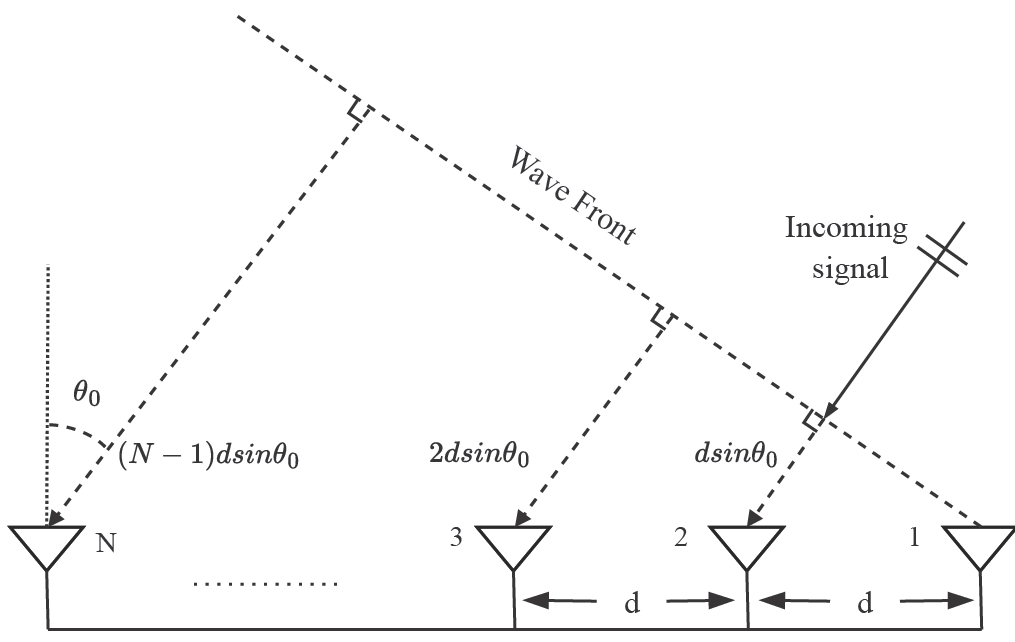}}
\caption{Plane wave impinging on a uniform linear array.}
\label{fig:IncomingSignal-example}
\end{figure}

A Uniform Linear Array (ULA), characterized by antennas equally spaced along a linear path, serves as the foundation for examining the impact of signal arrival angles and the spatial separation between antennas. The variation in propagation distance for a plane wave reaching each antenna element is a function of the arrival angle, denoted as $\theta$, and the inter-element spacing, $d$. This configuration, which is pivotal for understanding beamforming dynamics, is depicted in Fig. \ref{fig:IncomingSignal-example} \cite{torres2006implementation}.

As all antennas are uniformly distributed across the linear array, the varying path distance covered by the signal wave between consecutive antennas is given by $d \sin(\theta)$ when considering the angle of arrival $\theta$ perpendicular to the array. This difference results in a corresponding time delay:
\begin{equation}
\begin{aligned}
\tau(\theta)=\frac{dsin(\theta)}{c}
\end{aligned}
\end{equation}
where $c$ represents the speed at which the signal wave propagates. Consequently, the delay to the $n^{th}$ antenna concerning the first antenna of the array is given by:
\begin{equation}
\begin{aligned}
\tau(\theta)=(n-1)\frac{dsin(\theta)}{c}
\end{aligned}
\end{equation}
As each element within the linear array receives the wavefront signal with a specific time delay $\tau_n$, the retrieval of the signal from the desired direction can be achieved through linear processing. In this scenario, the filter for each channel $n$ corresponds to a time delay $\tau_n$ associated with that channel. Once all the spatial samples are temporally aligned, they can be summed to reconstruct the signal.

Assuming a propagating signal $x(t) = x_{0}(t)\cos(\omega_{c}t)$ arriving at an angle $\theta_{s}$ perpendicular to the ULA and modulated at a carrier frequency $\omega_{c}$, if the bandwidth of the baseband signal $x_{0}(t)$ is less than one percent of the carrier frequency, it is considered a narrowband beamformer \cite{ingle2005statisical}. In this context, the propagation delay between elements within the array can be approximated as a phase shift. Throughout this work, a narrowband beamformer will be assumed.
\begin{equation}
\begin{aligned}
x_{0}(t-\tau_{n})\approx x_{0}(t)
\end{aligned}
\end{equation}
Therefore, the signal at the $n^{th}$ antenna in the array can be expressed as:
\begin{equation}
\begin{aligned}
x_{n}(t)&=x_{0}(t)cos(\omega_{c}t-\theta_{n})=x_{0}(t)cos(\omega_{c}t-\omega_{c}\tau_{n})= \\ 
&=\Re(x_{0}(t)e^{-j\omega_{c}t}e^{-j\omega_{c}\tau_{n}})
\end{aligned}
\end{equation}
From this simplification, the delay lines associated with $\tau_{n}$ can be replaced with a phase shift $e^{-j\omega_{c}\tau_{n}}$.

\begin{figure}[t]
\centerline{\includegraphics[width=16pc]{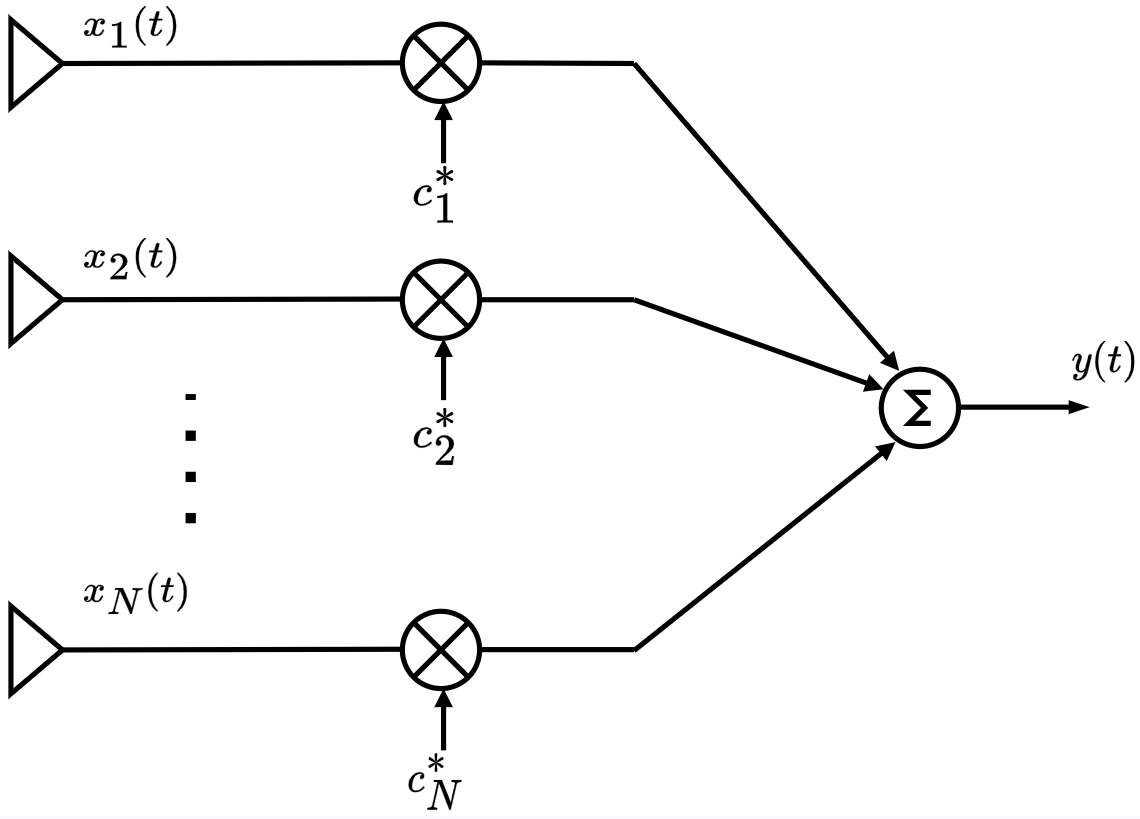}}
\caption{Beamforming operation.}
\label{fig:BeamformingOperation-example}
\end{figure}

The beamformer generates its output by forming a weighted combination of signals from the $N$ elements of the antenna array. This can be expressed as:
\begin{equation}
\begin{aligned}
y(t)=\sum_{n=1}^{N} c_{n}^{*}x_{n}(t)=\mathbf{c}^{H}\mathbf{x}(t)
\end{aligned}
\end{equation}
where $\mathbf{x}(t)$ is a vector containing the signals captured at every element within the array:
\begin{equation}
\begin{aligned}
\mathbf{x}(t)={\left[x_{1}(t)\ x_{2}(t)\ ...\ x_{N}(t)\right]}^T
\end{aligned}
\end{equation}
And $\mathbf{c}$ is the column vector of beamforming weights, named as the array response vector:
\begin{equation}
\begin{aligned}
\mathbf{c}&={\left[c_{1}\ c_{2}\ ...\ c_{N}\right]}^T={\left[e^{-j\omega_{c}\tau_{1}}\ e^{-j\omega_{c}\tau_{2}} \ ...\ e^{-j\omega_{c}\tau_{N}}\right]}^T=\\
&={\left[1\ e^{-j\omega_{c}\frac{dsin(\theta)}{c}} \ ... \ e^{-j\omega_{c}(N-1)\frac{dsin(\theta)}{c}}\right]}^T
\end{aligned}
\end{equation}
To achieve this, the beamforming operation linearly combines signals from all the antennas by applying specific weightings. This weighting process, illustrated schematically in Fig. \ref{fig:BeamformingOperation-example}, accentuates signals from a specific direction while simultaneously attenuating signals from different angles. Commonly referred to as 'electronic' steering, this beamforming operation assigns weights through electronic circuitry following signal reception or transmission to orient the array in the desired direction. This operation embodies a spatial filtering technique, analogous to temporal filtering, as it processes information across the elements within the antenna array to generate the desired response.

\subsection{Standard Digital Beamforming Receiver}
\begin{figure}[t]
\centerline{\includegraphics[width=20pc]{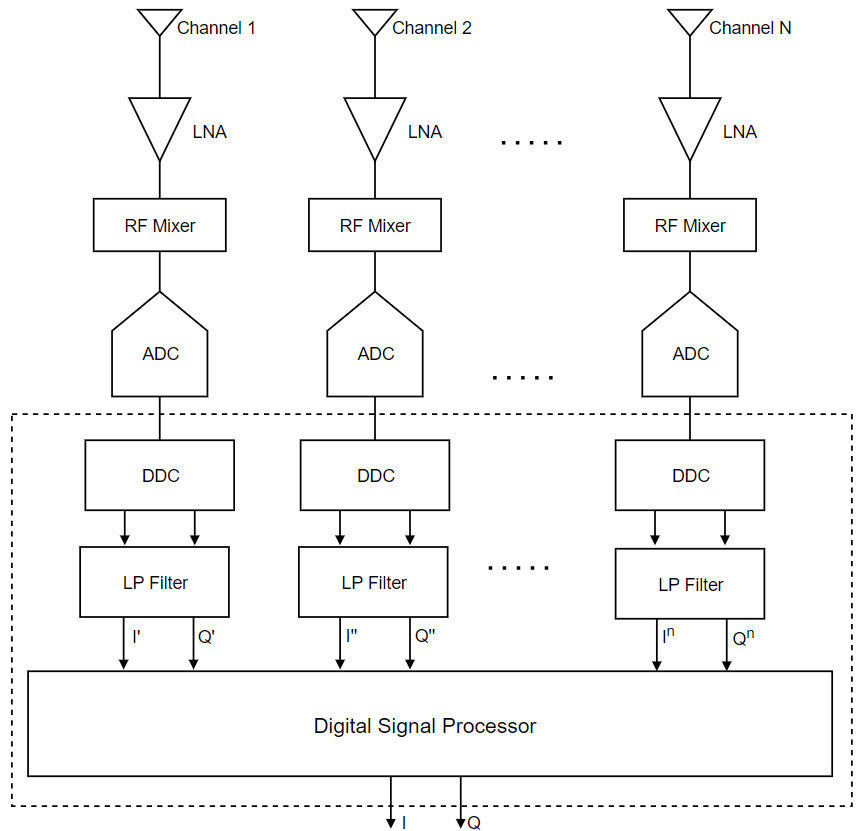}}
\caption{Standard Digital Beamforming Receiver.}
\label{fig:DBF-RX-example}
\end{figure}

The architecture of a conventional digital beamformer receiver is delineated by five principal components: the Radio Frequency (RF) Modulation Stage, Analog-to-Digital Converter (ADC), Digital Down Conversion (DDC) stage, Complex Weight Multiplication (CWM) stage, and the Summation stage. The culmination of this process—specifically the operations within the CWM and Summation stages—embodies the core of beamforming, effectively facilitating the linear combination of signals \cite{wirth2001radar}\cite{litva1996digital}\cite{torres2006implementation}\cite{LowCostDBFRX}. Fig. \ref{fig:DBF-RX-example} provides a schematic overview of this traditional beamforming receiver architecture.

Commencing the signal processing journey, the RF Modulation stage serves as the primary interface with the antenna channel. Its chief function is to downconvert the carrier frequency to a more manageable intermediate frequency (IF) via an RF Translator or mixer, usually located proximate to a Low Noise Amplifier (LNA) to minimize the noise figure and amplify the Signal-to-Noise Ratio (SNR). This modulation stage is critical, particularly because the frequencies of incoming signals often surpass the operational speed of available ADCs, necessitating analog signal modulation to transpose the signal's frequency components into a band that is within the ADC's capture range. Nevertheless, advancements in high-speed ADC technology now permit the direct acquisition of RF signals through undersampling techniques, potentially obviating the need for a separate RF modulation stage \cite{jayamohan2015not}.

Following RF modulation, the signal encounters the ADC stage, where the analog waveform is transformed into a digital format, ready for subsequent digital beamforming operations. This stage encompasses the critical functions of sampling, quantizing, and encoding the analog signal into its digital counterpart.

The DDC stage marks the commencement of digital signal processing within the DBF Receiver, immediately following the ADC. In this stage, the digitized signal is processed to generate the in-phase (I) and quadrature (Q) components, which are then translated into the baseband frequency range. A low-pass filter is subsequently applied to both I and Q signals to remove any image frequency components, effectively lowering the data rate for the digital platform. This reduction facilitates the efficient execution of multiplications and additions in the later stages of signal processing. To streamline the DDC process, adopting a sampling frequency that positions the sampled spectra exactly at $\pm f_{s}/4$ is recommended \cite{lyons1997understanding}. This strategy eliminates the necessity for physical mixers or multipliers for down-conversion, instead utilizing a sign inversion process on selected samples—a technique further elaborated upon in later discussions.

The final components of the DBF Receiver are the CWM and Summation stages, which together constitute the core of the Digital Signal Processor, as illustrated in Fig. \ref{fig:DBF-RX-example}. These stages represent the zenith of computational demand within the receiver. The CWM stage is responsible for processing inputs such as the in-phase and quadrature baseband signals, alongside the magnitude and phase of the complex weights. The Summation stage, on the other hand, aggregates all I and Q baseband signals emanating from the antenna channels, completing the beamforming process.

This paper utilizes the standard digital beamforming receiver design as a benchmark to evaluate the performance and resource efficiency of the proposed architecture when implemented on an FPGA platform.

\subsection{Under-sampling}
ADCs play a crucial role in the digital beamforming receiver architecture by sampling incoming RF signals and converting them into digital words. According to the Nyquist theorem, a sampling rate greater than twice the highest frequency within the signal's bandwidth is theoretically sufficient for the unique reconstruction of the signal \cite{pelgrom2013analog}. This principle establishes a critical relationship between the signal's bandwidth (BW) and the minimum sampling rate ($f_{s}$):
\begin{equation}
\begin{aligned}
f_{s} > 2BW
\end{aligned}
\end{equation}

The Nyquist theorem's flexibility allows for the sampling of an incoming RF signal at a higher frequency band, without starting from `0 Hz'. This capability facilitates the placement of the signal's band of interest at frequencies above or even beyond the sampling rate. Known as `under-sampling', `down-sampling', `bandpass sampling', or `sampling translation', this technique integrates digitization and frequency translation of the RF signal in a singular operation. Each sampling instance inherently replicates the spectrum, a fundamental aspect in digital signal processing that underpins demodulation in radio-communication systems. Under-sampling offers significant benefits, including reduced power consumption, cost savings, and minimized board space, by negating the necessity for an additional mixer.

\begin{figure}[t]
\centerline{\includegraphics[width=20
pc]{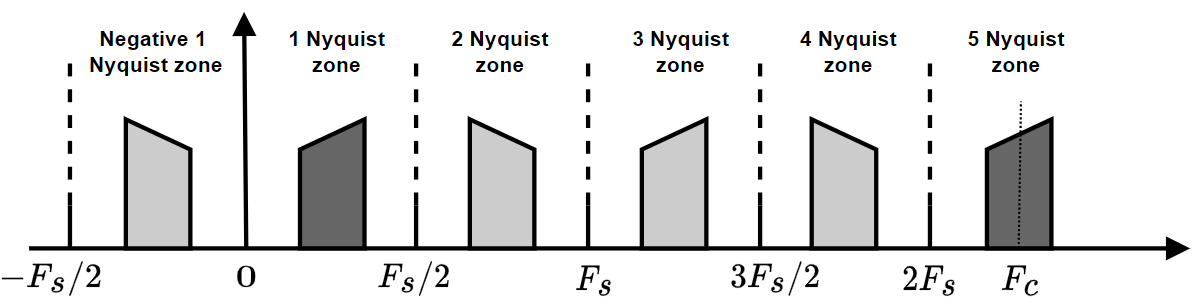}}
\caption{Under-sampling spectrum.}
\label{fig:Under-sampling-example}
\end{figure}

For effective under-sampling, adherence to the Nyquist criterion and meticulous frequency planning are essential, particularly within designated Nyquist zones. These zones are segments of the frequency spectrum spaced at intervals of $f_{s}/2$. Each zone can host either an exact or a mirrored replica of the desired signal's spectrum, with odd zones presenting direct replicas and even zones showing mirrored versions \cite{pearson2011high}. Fig. \ref{fig:Under-sampling-example} illustrates how the original signal's spectrum, located in the fifth Nyquist zone, undergoes this under-sampling process.

The formula for determining the under-sampling frequency rate, to ensure the signal's spectrum is correctly positioned within a specific Nyquist zone, is as follows \cite{rouphael2009rf}:
\begin{equation}
\begin{aligned}
\frac{2f_{c}+BW}{2n+1}&\leq f_{s}\leq \frac{f_{c}- BW/2}{n}; \\
0&\leq n\leq \frac{f_{c}- BW/2}{2BW} 
\end{aligned}
\end{equation}
For inverted spectral placement of the signal, the sampling frequency range is given by the following expression:
\begin{equation}
\begin{aligned}
\frac{f_{c}+BW/2}{n}&\leq f_{s}\leq \frac{2f_{c}-BW}{2n-1}; \\
1&\leq n\leq \frac{f_{c}+BW/2}{2BW} 
\end{aligned}
\end{equation}

\section{HARDWARE IMPLEMENTATION}

\begin{figure}[t]
\centerline{\includegraphics[width=20pc]{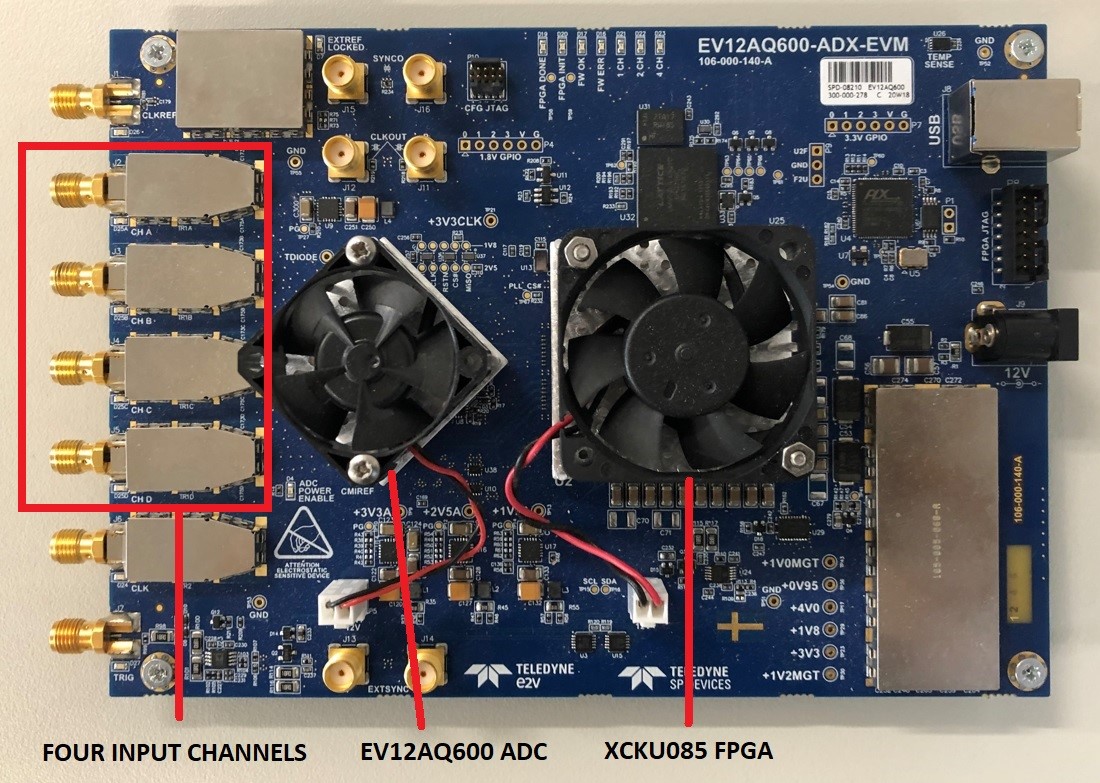}}
\caption{EV12AQ60X-ADX-EVM evaluation board.}
\label{fig:EV12AQ60X-ADX-EVM}
\end{figure}

In the development and validation of the proposed DBF architecture for spaceborne applications, the EV12AQ60X-ADX-EVM evaluation board by Teledyne e2V Semiconductors was employed (refer to Fig. \ref{fig:EV12AQ60X-ADX-EVM}). This board is equipped with the space-qualified, high-speed ADC EV12AQ600, which boasts a quad-channel, 12-bit configuration, along with the high-performance Field-Programmable Gate Array (FPGA) from Xilinx's Kintex UltraScale series, model XCKU085. The evaluation module (EVM) features four SMA connectors on its edge, each corresponding to an analog input channel designed for RF signal acquisition. In addition to these, the board is furnished with additional SMA connectors to support a range of functionalities, including but not limited to, external clock reference, external trigger inputs, and external clock inputs. The Phase-Locked Loop (PLL) component LMX2592, integrated onto the board, plays a crucial role in clock management and initialization processes for both the ADC/FPGA interface and the FPGA's data pathway \cite{LMX2592}.

The EVM is designed to operate in three distinct modes, each determined by the configuration of its interleaving functionality. These modes include: a one-channel mode, where a single analog input channel is directed to all four cores of the ADC; a two-channel mode, wherein one analog input channel is allocated to two ADC cores, and a second analog input channel is distributed to the remaining two cores; and a four-channel mode, which assigns each of the four analog input channels to an individual ADC core. For the purposes of this study, the EVM was exclusively configured to function in the four-channel mode. The board is equipped with a series of internal registers, enabling adjustments to various parameters including the ADC's operational characteristics, the LMX2592 PLL settings, channel mode selection, clock configurations, and data acquisition parameters.

\begin{figure}[t]
\centerline{\includegraphics[width=20pc]{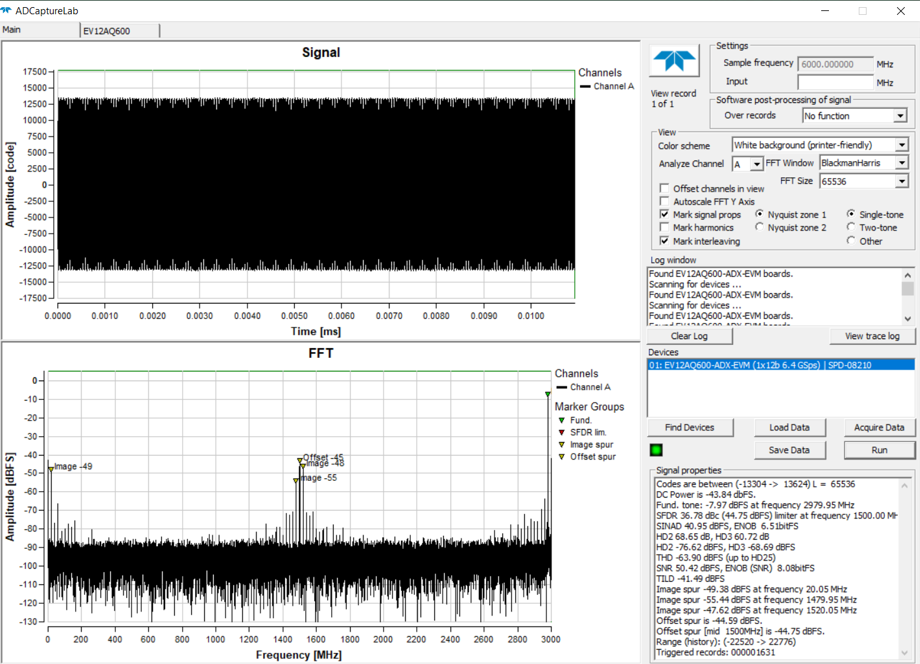}}
\caption{ADCaptureLab screenshot main window.}
\label{fig:ADCaptureLab screenshot}
\end{figure}

The ADCaptureLab, a graphical application developed by Teledyne e2V Semiconductors, serves as a comprehensive tool for managing the configuration of the EVM, initiating data acquisition processes, and facilitating the visualization of captured data within both time and frequency domains. This application is instrumental in computing key performance indicators, including SNR, Signal-to-Noise and Distortion Ratio (SNDR), Spurious Free Dynamic Range (SFDR), and Total Harmonic Distortion (THD) \cite{EVM}, as illustrated in the graphical representation provided in Fig. \ref{fig:ADCaptureLab screenshot}. Within the scope of this research, the ADCaptureLab software has been pivotal for the ADC characterization process.

For additional operations concerning the EVM, such as hardware configuration adjustments, channel mode selection, and modifications to the registers of the ADC, FPGA, and PLL, a series of Python scripts were developed and utilized. Moreover, the digital signal processing functionalities, including but not limited to adders, multipliers, complex down-conversion, and Finite Impulse Response (FIR) low-pass filtering, which form the backbone of the digital beamforming receiver, were meticulously designed and realized on the XCKU085 FPGA. The implementation was conducted using VHDL as the Hardware Description Language (HDL) through the employment of the Xilinx Vivado 2020.2 (64-bit) software suite.

The outcomes of the operational deployment, which encompasses the baseband filtered beam, are conserved within a Random Access Memory (RAM) module situated inside the FPGA, designated as the DBF debug module. This module has been specifically engineered to enable the transmission of the stored samples from the RAM to a personal computer (PC) via Universal Asynchronous Receiver-Transmitter (UART) protocol, thereby simplifying the process of visualizing the reconstructed baseband signals. Leveraging the identical evaluation board and development framework, a conventional digital beamformer receiver architecture, as delineated in the preceding section, was also materialized. This concurrent implementation permits an exhaustive comparative analysis of both designs, focusing on crucial metrics such as resource component utilization, power consumption, and temporal efficiency.

\subsection{EV12AQ600 Analog to Digital Converter}
The ADC device selected for the under-sampling of RF signals in this study is the EV12AQ600, supplied by Teledyne e2v Semiconductors and integrated within the EV12AQ60X-ADX-EVM evaluation board. This device is meticulously engineered to meet the stringent requirements of the European Space Components Coordination (ESCC) and QML-Y space specifications. The EV12AQ600, a quad-channel, 12-bit ADC, demonstrates a remarkable capability for sampling rates up to 1.6 Giga Samples per Second (GSps). It is designed to support versatile multi-mode operation, allowing the interleaving of its four independent cores to enhance the sampling rate, achieving up to 6.4 GSps in a single-channel mode. For the purposes of this investigation, the EV12AQ600 was configured to operate in a 4-channel mode. In this configuration, each of the four cores independently samples signals from four distinct antenna inputs within an array, each at a rate of 1.6 GSps \cite{EV12AQ600}.

\begin{figure}[t]
\centerline{\includegraphics[width=20pc]{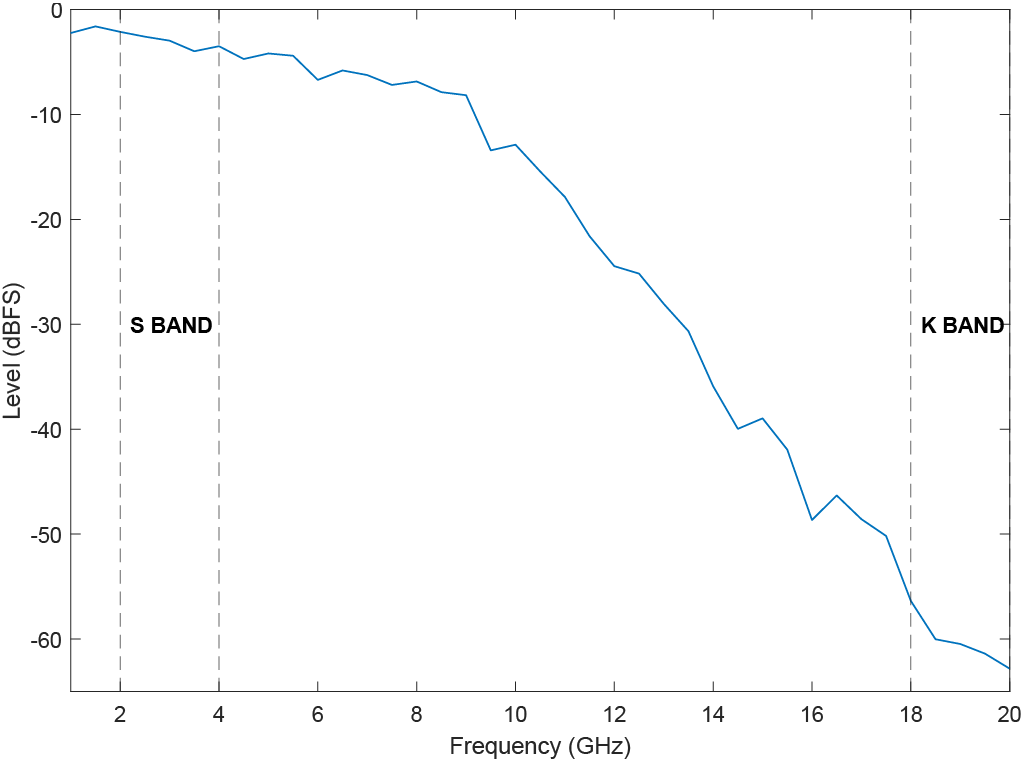}}
\caption{ADC Analog equivalent bandwidth.}
\label{fig:ADC Analog equivalent bandwidth}
\end{figure}

\begin{figure}[t]
\centerline{\includegraphics[width=20pc]{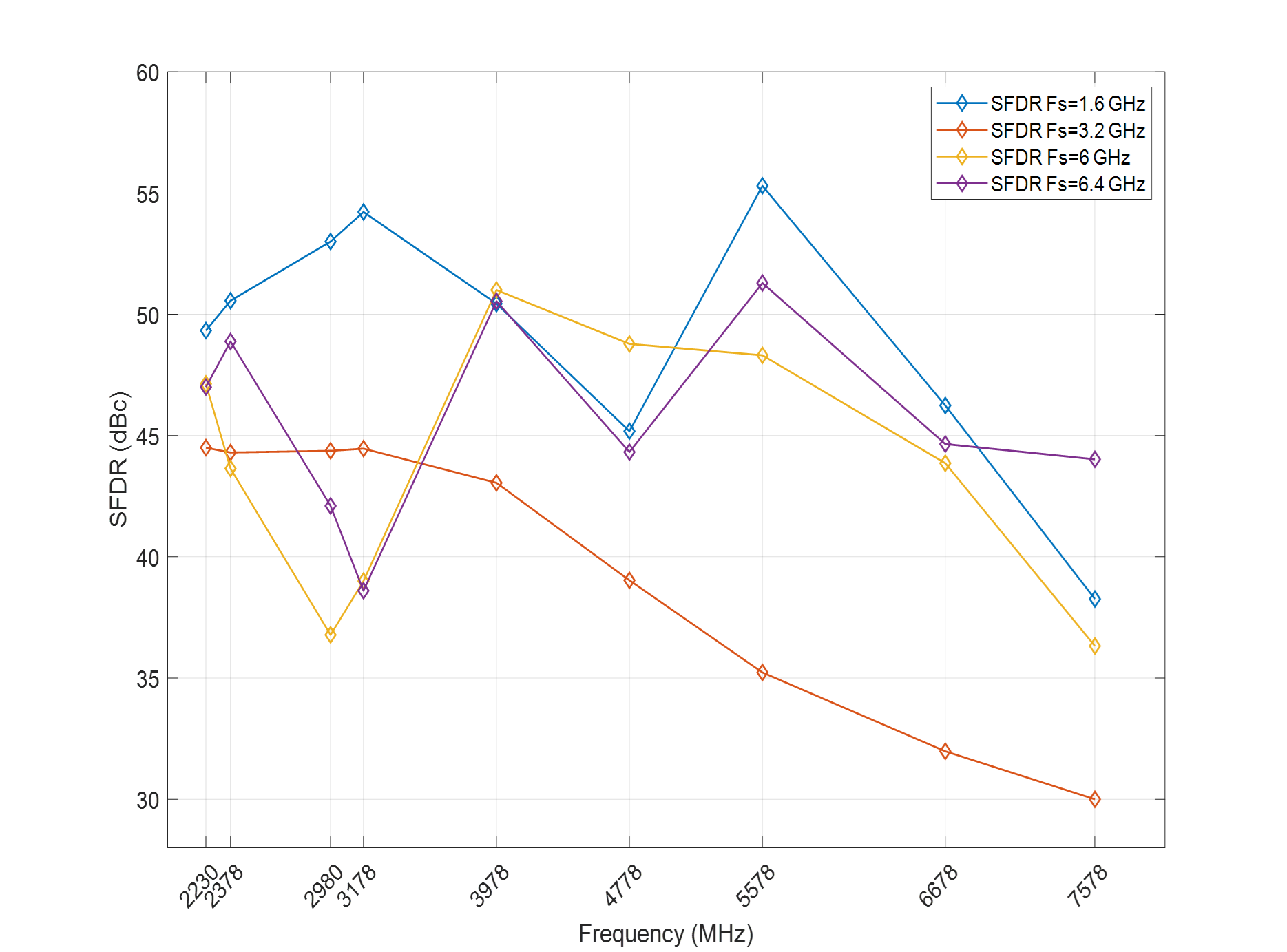}}
\caption{ADC SFDR Performance metrics.}
\label{fig:ADC SFDR Performance metrics}
\end{figure}

The characterization of the EV12AQ600 ADC, employed for undersampling processes on RF signals, was conducted by measuring its response to input tones spaced at 0.5 GHz across a range from 1 to 20 GHz. These tests were performed under laboratory conditions using the R\&S SMR20 Signal Generator and the Agilent N9010A EXA Signal Analyzer. The configuration of the EV12AQ600 and the collection of data from the digitized signals were facilitated through the ADCaptureLab software application, provided by Teledyne, and connected to a PC via USB cable. This software enables the capture of sampled data, visualization of the signal waveform in the time domain, plotting of its Fast Fourier Transform (FFT), and obtaining crucial spectral performance parameters. It also allows for the configuration of the EV12AQ600 according to the channel operating mode, frequency rate, etc. The frequency response of the ADC, measured for various injected input tones, is presented in Fig. \ref{fig:ADC Analog equivalent bandwidth}. The obtained analog equivalent bandwidth reveals a marked attenuation of the signal starting around 5.5 GHz. This indicates that the EV12AQ600 effectively processes radio frequency signals in the S band and part of the C band up to 5.5 GHz, while signals beyond this frequency experience significant attenuation. Fig. \ref{fig:ADC SFDR Performance metrics} illustrates the performance of the EV12AQ600 ADC in terms of its SFDR values at various sampling rates for specific input RF frequencies.

\subsection{ESIstream High-Speed Serial Link Protocol}
The digital data produced by the EV12AQ600 ADC is conveyed to the FPGA via a serial communication link, leveraging the ESIstream protocol \cite{ESIstreamProtocol}. This open-license protocol, devised by Teledyne e2v Semiconductor, is tailored for serial communications between high-speed data converters and FPGAs, enhancing interoperability and efficiency. The ESIstream architecture encompasses the EV12AQ600 as the transmitter (TX), the FPGA as the receiver (RX), and utilizes eight lanes to transmit data frames from each ADC core, with data being output across two serial links per core, supplemented by a synchronization signal to facilitate the initiation of communication.

Employing a 14b/16b encoding scheme, the ESIstream protocol achieves an efficiency of 87.5\%, presenting a notable improvement over the 80\% efficiency typically observed with 8b/10b encoding. Key attributes of the protocol include its support for deterministic communication with minimal link latency, capability for a lane rate up to 12.8 Gbps, synchronization across multiple serial lanes, and the facilitation of multi-device synchronization. This makes it particularly suitable for configurations involving numerous transmission and reception modules, ensuring robust and efficient data exchange.

For the configuration and deployment of the TX and RX modules utilizing the ESIstream protocol, Teledyne provides a specialized IP core compatible with the Xilinx Vivado design suite \cite{ESIstreamIPCore}. This IP core is tailored specifically for use with the targeted FPGA platform, ensuring seamless integration and optimal performance. Presently, the market also offers ESIstream IP cores designed for compatibility with several leading space-grade Serializer/Deserializer (SerDes) technologies, including the Xilinx XCKU60, Xilinx Virtex 5QV, Microsemi Polar Fire, and Microsemi RTG4, catering to a wide range of application requirements.

Upon implementation within the FPGA, the ESIstream IP core's RX module constructs a data framework that delivers 8 parallel samples for each channel, with each sample being 12 bits in length, operating at a clock frequency of 200 MHz. This configuration is meticulously aligned with the output specifications of the EV12AQ600 ADC, which performs sampling across each input channel at a rate of 1.6 GSps, producing 12-bit data words. This synchronicity ensures that the high-speed data transmission from the ADC to the FPGA is managed efficiently, with precision and minimal latency.
To facilitate the evaluation of the serial communication link and ensure the precise transmission of data, Teledyne offers a comprehensive Xilinx Vivado project example written in VHDL, designed specifically for interfacing the EV12AQ600 ADC with an FPGA. This example project includes a diagnostic module that scrutinizes the integrity of the decoded frames received, enabling the detection of any data or clock bit errors within the frames. The diagnostic outcomes, including the error status of the received data frames and various ESIstream control signal states, are seamlessly integrated with Integrated Logic Analyzers (ILA) for in-depth analysis within the Vivado hardware manager, utilizing the JTAG port for connectivity.

During the evaluation of the serial communication efficacy between the EV12AQ600 and the FPGA, an anomaly was detected through the ILA in the data frame alignment for two of the four channels within the RX ESIstream module IP core. Specifically, the frames received through these channels were misaligned, leading to a distortion of the conveyed information. This misalignment issue was promptly communicated to Teledyne, which responded by issuing an updated version of the RX module IP core. This revised IP core successfully rectified the alignment issue, thereby restoring the integrity of the data transmission process.
\subsection{Xilinx Kintex UltraScale XCKU085 FPGA}
The EV12AQ60X-ADX-EVM evaluation board integrates a high-performance FPGA, the Xilinx Kintex UltraScale XCKU085 device. This FPGA leverages advanced technologies, including monolithic structures and next-generation stacked silicon interconnects, to meet the demands for elevated digital signal processing capabilities and support high-speed serial data communication. These features render it ideal for the requirements specified in this work \cite{UltraScaleOverview}. Equipped with 48 GTH high-speed serial transceivers, the FPGA utilizes 8 of these for the reception of data sampled by the EV12AQ600 ADC, which is then transmitted to the FPGA using the ESIstream protocol.

The GTH transceivers are designed to support a wide range of line rates, from 500 Mb/s up to 16.375 Gb/s, with the effective data capacity per lane reaching up to 12.5 Gbps. These transceivers are organized into quartets and employ the CML interface standard, ensuring each transceiver’s throughput is more than adequate for the data rates produced by the ADC. The deployment of these transceivers within the hardware design is streamlined by the UltraScale FPGAs Transceivers Wizard, a utility provided by Xilinx. This wizard assists in the automatic creation of the XDC constraint file, which can be further customized to adjust operating conditions and placement specifics tailored to the project's needs.
The digital signal processing tasks within the FPGA are executed utilizing specialized units known as DSP slices, specifically the DSP48E2 blocks, as delineated by the manufacturer \cite{UltraScaleDSP}. The strategic deployment of these DSP resources significantly enhances the efficiency of processing operations, optimizing both speed and power consumption. Thus, a critical aspect of the hardware design phase involves the careful allocation of these DSP slices to meet the processing demands effectively.

While the FPGA model used in this study, the Xilinx Kintex UltraScale XCKU085, is not classified as space-qualified, its counterpart, the Xilinx Kintex UltraScale XQRKU060, is designed for space applications. The XQRKU060 shares a similar architecture and DSP resources with the XCKU085, though it possesses a slightly reduced count of these resources. Nevertheless, the XQRKU060's capabilities are deemed sufficient for the implementation requirements of this project.

The hardware design for the Digital Beamforming Receiver on the XCKU085 FPGA has been finalized using the Xilinx Vivado 2020.2 (64-bit) design suite, employing VHDL for development. The programming of the FPGA is facilitated through a JTAG connection.

\section{DIGITAL BEAMFORMING RECEIVER PROPOSED}
\begin{figure}[t]
\centerline{\includegraphics[width=14pc]{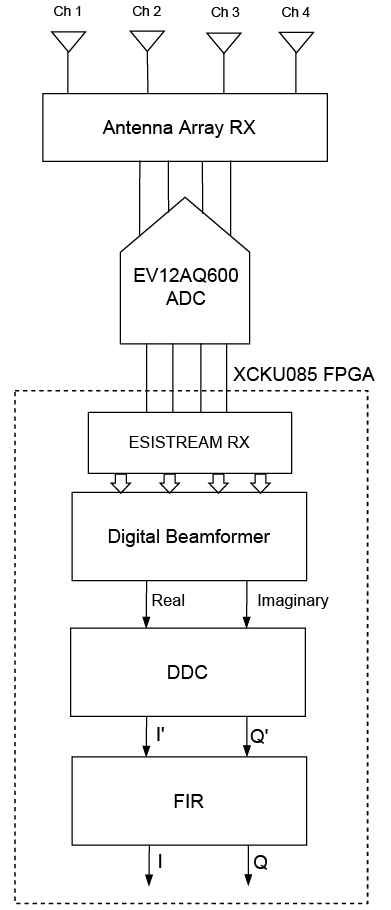}}
\caption{Digital Beamforming Receiver design.}
\label{fig:Digital Beamforming Receiver design}
\end{figure}

Leveraging the advanced hardware framework outlined in the previous section, which includes the high-speed ADC and the high-performance FPGA, this study introduces a sophisticated digital beamformer receiver design. This design, featuring four input channels, is detailed in Fig. \ref{fig:Digital Beamforming Receiver design}.
The Teledyne e2v's EV12AQ600 ADC plays a pivotal role in this architecture, capturing RF signals from a four-antenna array receiver and converting these into 12-bit digitized samples at a significant rate of 1.6 Gbps. This ADC employs an under-sampling technique to shift the RF signals to an IF, facilitating a more manageable frequency range for processing. These digitized samples are then seamlessly transmitted to the XCKU085 FPGA via the ESIstream protocol over a high-speed serial link (HSSL), ensuring efficient and rapid data transfer.

As illustrated in Fig. \ref{fig:Digital Beamforming Receiver design}, upon reception by the FPGA through its high-speed serial transceivers, the four channels are subject to a linear combination process. This critical step involves the multiplication of each channel by a designated complex weight, followed by the aggregation of all channels, culminating in a composite output channel with both real and imaginary components.

The DSP48E2 blocks within the FPGA are tasked with executing the beamforming operations, significantly enhancing the efficiency of digital signal processing tasks. These operations benefit from reduced power consumption and the acceleration of computational elements such as adders and multipliers, which are essential for executing multiply-accumulate (MAC) operations. The XCKU085 FPGA adeptly handles the beamforming digital signal processing tasks at the data rates facilitated by the ESIstream protocol, prior to the signal's downconversion to baseband. A notable advantage of this design is its efficiency post-linear combination, where only a singular channel, composed of real and imaginary components, undergoes processing through DDC and a low-pass FIR filter.

The integration of additional antenna channels and ADC devices into the system accentuates the resource efficiency achieved through the digital beamformer design. Despite the expansion, the output from the digital beamforming stage is maintained as a single channel due to the linear combination process, thereby yielding significant savings in system resources. Given that the sampling frequency ($F_{s}$) from the ADC is set to four times the intermediate frequency ($F_{c}$), the DDC stage employs a unique processing technique on select samples, which obviates the need for multipliers or RF mixers traditionally used for down-converting the passband signal.

Adopting a sampling frequency that is quadruple the intermediate frequency ($F_{s}=4F_{c}$) brings forth multiple benefits. A primary advantage is the simplification of the mixing process in I-Q demodulation. In this setup, the cosine and sine waveforms generated by the numeric oscillator are sampled at quadruple their inherent frequencies. This alteration transforms the mixing process into straightforward multiplication by the sequences ${1, 0, -1, 0}$ for the cosine component and ${0, 1, 0, -1}$ for the sine component, effectively executed using switching and sign inversion techniques, as depicted in Fig. \ref{fig:Complex Down-conversion schematic}. This strategic approach confines the intermediate frequency to 400 MHz post under-sampling, correlating with the ADC's sampling rate of 1.6 GHz.

An illustration of an input signal, characterized by an intermediate frequency (IF) of 400 MHz, is depicted in Fig. \ref{fig:Frequency modulated input signal}. To mitigate the presence of image frequency components within both the in-phase and quadrature output signals, a FIR low-pass filter is employed. This filter features 64 taps, with each tap being 10 bits in length, ensuring effective suppression of unwanted frequency components and enhancing signal fidelity.
\subsection{Digital Beamformer}
\begin{figure}[t]
\centerline{\includegraphics[width=20pc]{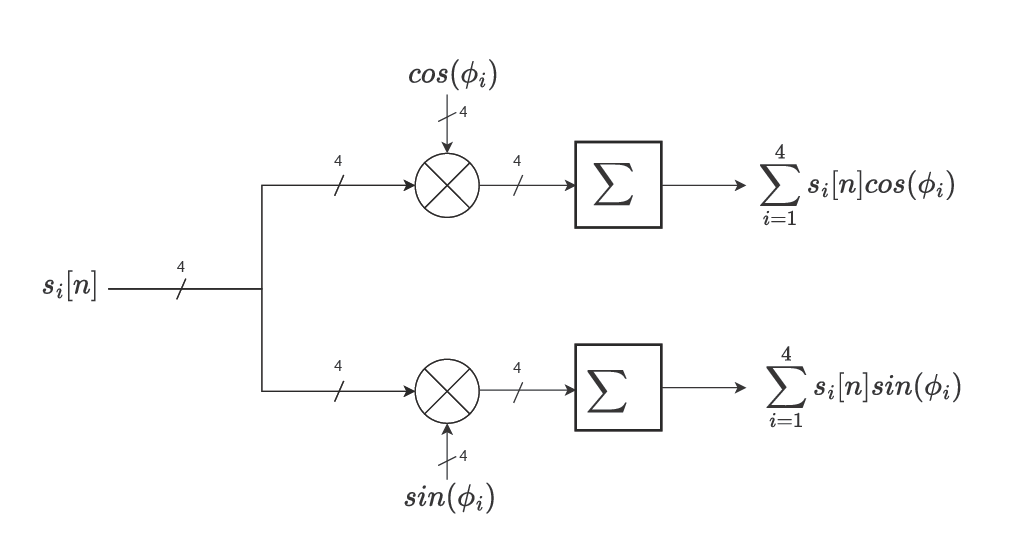}}
\caption{Digital Beamformer schematic.}
\label{fig:Digital Beamformer schematic}
\end{figure}
At this stage, the digital IF signal for each channel, streamed to the XCKU085 FPGA via the ESIstream protocol, undergoes a critical transformation. Each channel's signal is multiplied by corresponding complex weight values, which are then summed together. This process effectively creates a partial passband beam across the four channels. The complex weight values, assigned a 12-bit length for each channel, are stored in dedicated registers within the FPGA. The hardware logic is designed to fetch these values for the purpose of executing multiplications with the incoming data streams.

The data flow emerging from the ESIstream protocol for each channel is characterized by 8 parallel samples, each with a 12-bit resolution, and is clocked at a frequency of 200 MHz. Following the process of linear combination within the digital beamformer, the resultant samples are similarly dispatched at a 200 MHz clock frequency, maintaining the parallel output of 8 samples. These samples are subject to truncation via a configurable window mechanism, which facilitates adjustment according to the dynamic range, thereby enabling the selection of a specific sample subset. The positioning of this window is governed by a parameter stored within a FPGA register, dictating the bit count to be read starting from the least significant bit. By default, this window prioritizes the most significant bits for consideration. Consequently, the output samples from the beamforming stage are truncated to a 20-bit length.

This module produces a complex output, as described by the following mathematical expression:
\begin{equation}
\begin{aligned}
y(t)=&\sum_{i=1}^{4} s_{i}[n]e^{j\phi_{i}}= \\ 
=&\sum_{i=1}^{4} s_{i}[n]cos(\phi_{i}) + j\sum_{i=1}^{4} s_{i}[n]sin(\phi_{i})
\end{aligned}
\end{equation}
where $s_{i}[n]$ are the four IF input signals digitalized by the EV12AQ600 ADC device at 1.6 GHz and $\phi_{i}$ are the beamforming phase delays assigned to every channel. The schematic is shown in Fig. \ref{fig:Digital Beamformer schematic}.
In the development of the Digital Beamformer Receiver, the design strategically employs 64 DSP slices from the XCKU085 FPGA. This allocation corresponds directly to the necessity for 64 multiplication and 64 addition operations integral to the beamforming process. This efficient utilization of FPGA resources highlights a significant reduction when compared to traditional digital beamformer configurations, a comparison that will be elaborated upon subsequently. Throughout the design phase, the allocation of DSP slices was maximized to fully exploit the technological advantages offered by the FPGA in terms of performance enhancement.

The synthesis tool employed during the design phase automatically infers the DSP48E2 slices for arithmetic operations from the VHDL code. Nonetheless, for designers seeking more granular control over DSP resources, Xilinx provides the option to explicitly instantiate the DSP48E2 primitive. This approach facilitates direct access to the DSP slices, offering an avenue for optimized resource management and potentially, enhanced system performance \cite{UltraScaleDSP}.

\subsection{Complex Down-conversion}
In the digital beamformer receiver, the complex down-conversion module plays a pivotal role in translating the digitized signal's discrete spectrum to be centered around zero Hz. By setting the intermediate frequency ($f_{IF}$) to a quarter of the sampling frequency ($f_{s}/4$), the outputs of the cosine and sine oscillators generate repetitive four-element sequences, specifically $cos(\pi n/2) = 1, 0, -1, 0$, and $-sin(\pi n/2) = 0, -1, 0, 1$ respectively \cite{7080413}. This approach obviates the need for physical mixers or multipliers for down-converting the spectra to baseband. Instead, the process is simplified to applying a sign inversion to select samples.

\begin{figure}[t]
\centerline{\includegraphics[width=20pc]{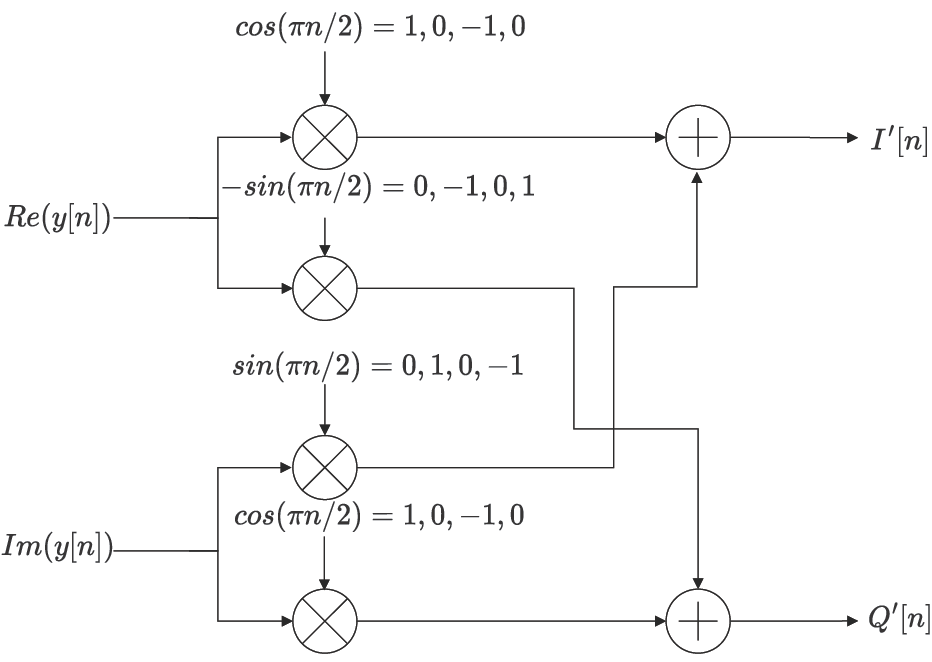}}
\caption{Complex Down-conversion schematic.}
\label{fig:Complex Down-conversion schematic}
\end{figure}

With the sampling frequency ($f_{s}$) established at 1.6 GHz, the intermediate frequency is accordingly set to 400 MHz. The output from this module is characterized by eight parallel samples, each extending to 20 bits in length, for both the in-phase and quadrature components, and is clocked at a frequency of 200 MHz. The schematic configuration of this complex down-conversion process is depicted in Fig. \ref{fig:Complex Down-conversion schematic}.
The mathematical expression of the in-phase and quadrature outputs from this module is the following:
\begin{equation}
\begin{aligned}
I^{'}[n]=\Re(y[n])cos\left(\frac{\pi n}{2}\right) + \Im(y[n])sin\left(\frac{\pi n}{2}\right)
\end{aligned}
\end{equation}
\begin{equation}
\begin{aligned}
Q^{'}[n]=-\Re(y[n])sin\left(\frac{\pi n}{2}\right) + \Im(y[n])cos\left(\frac{\pi n}{2}\right)
\end{aligned}
\end{equation}
This configuration of the complex down-conversion module does not require any DSP components, but only some minor LUT logic and CLB components.

\subsection{FIR Filter}
\begin{figure}[t]
\centerline{\includegraphics[width=20pc]{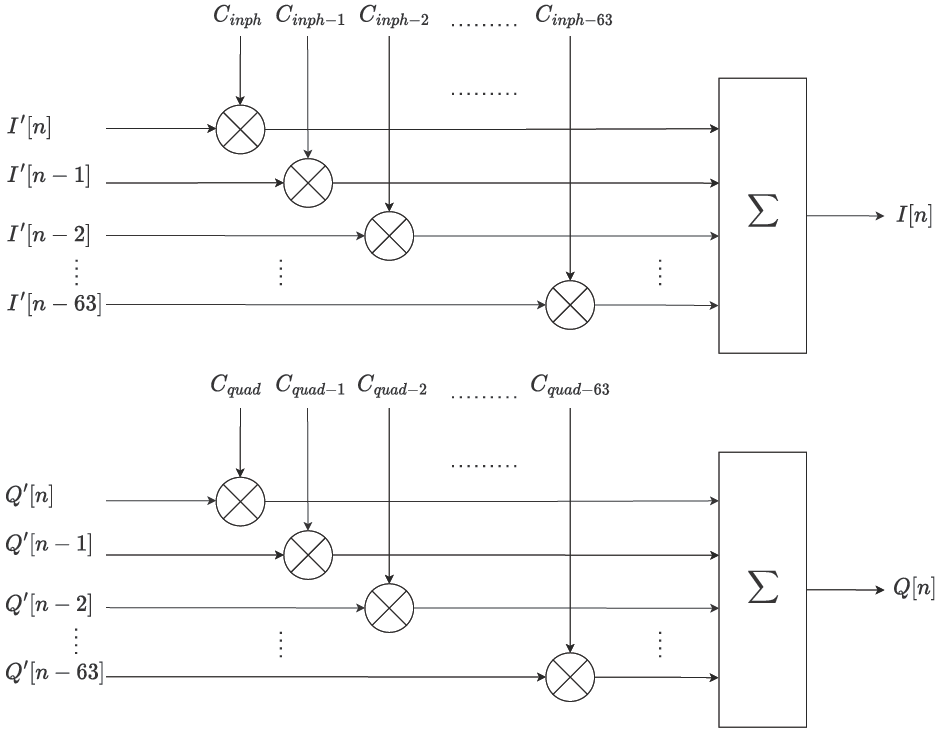}}
\caption{FIR filter schematic.}
\label{fig:FIR filter schematic}
\end{figure}

The module under discussion performs critical low-pass FIR filtering, effectively removing unwanted spectral components from both the in-phase and quadrature demodulated beams. It utilizes a configuration of 64 tap coefficients, each with a precision of 10 bits, applied identically to signals in both the in-phase and quadrature phases. These coefficients are securely stored within the FPGA's registers. The design and operational flow of this FIR filtering process are illustrated in Fig. \ref{fig:FIR filter schematic}. Post-filtering, the output from this module is delivered as 8 parallel samples, with each sample being 36 bits in length, for both the in-phase $I[n]$ and quadrature $Q[n]$ streams, functioning at a clock frequency of 200 MHz.

Implementing the FIR low-pass filter within this module requires the allocation of 1358 DSP slices from the XCKU085 FPGA, incorporating 1024 multipliers and 1008 adders. Although this stage is the most resource-intensive in terms of digital signal processing tasks, its application to a singularly linearly combined channel means that the resource demand remains consistent, irrespective of the number of RF input channels integrated into the array.

\subsection{Resource Utilization}
The resource utilization for every one of the stages described in the digital beamformer receiver for the XCKU085 FPGA and its total number is shown in Table \ref{tab:Resource utilization per stage}.
\begin{table}[htbp]
  \centering
  \caption{RESOURCE UTILIZATION PER STAGE}
    \resizebox{\columnwidth}{!}{
    \begin{tabular}{p{5.2em}|c|c|c|c|c}
    \multicolumn{1}{l|}{\textbf{Entity}} & \textbf{LUTs} & \textbf{Registers} & \textbf{CARRY8} & \textbf{CLB} & \textbf{DSPs} \\
    \hline
    Digital \newline{}Beamforming & \multirow{2}{*}{0}     & \multirow{2}{*}{0}     & \multirow{2}{*}{0}     & \multirow{2}{*}{0}     & \multirow{2}{*}{64} \\
    \hline
    Down-conversion & \multirow{2}{*}{152}   & \multirow{2}{*}{320}   & \multirow{2}{*}{24}    & \multirow{2}{*}{99}    & \multirow{2}{*}{0} \\
    \hline
    FIR Filter & 0     & 2200  & 0     & 653   & 1358 \\
    \hline
    Total & 152   & 2520  & 24    & 752   & 1422 \\
    \end{tabular}%
    }
  \label{tab:Resource utilization per stage}%
\end{table}%

As illustrated in Table \ref{tab:Resource utilization per stage}, the FIR filter stage is the most resource-intensive in this digital beamformer receiver design, consuming a significant portion of the DSP slices available in the XCKU085 FPGA. This stage, along with the digital beamformer, employs the FPGA's DSP slices for all its multiplier and adder operations, substantially enhancing the efficiency of digital signal processing tasks.

A distinctive aspect of this architecture is its scalability in terms of resource allocation. Introducing additional input channels to the Digital Beamformer does not impose additional resource demands on the Complex Down-conversion and FIR filter modules. These latter stages exhibit stable resource consumption and performance metrics, regardless of the number of RF input channels. For example, expanding the system to accommodate 16 input channels would necessitate a fourfold increase in resources for the Digital Beamformer stage—correlating with the quadrupled channel count—while the Complex Down-conversion and FIR filter stages would continue to operate with the resource allocation designated for four input channels.

Table \ref{tab:RESOURCE UTILIZATION PERCENTAGE} details the comprehensive component count and their respective utilization percentages within the XCKU085 FPGA, inclusive of the ESIstream receiver IP core, as derived from the Xilinx Vivado 2020.2 software suite.

\begin{table}[htbp]
  \centering
  \caption{RESOURCE UTILIZATION PERCENTAGE}
    \resizebox{\columnwidth}{!}{
    \begin{tabular}{l|c|c|c}
    \textbf{Resource} & \textbf{Utilization} & \textbf{Available} & \textbf{Utilization \%} \\
    \hline
    \textbf{LUT} & 6954  & 497520 & 1.4 \\
    \hline
    \textbf{LUTRAM} & 8     & 267840 & 0 \\
    \hline
    \textbf{FF} & 8061  & 995040 & 0.81 \\
    \hline
    \textbf{BRAM} & 93    & 1620  & 5.74 \\
    \hline
    \textbf{DSP} & 1422  & 4100  & 34.68 \\
    \hline
    \textbf{IO} & 55    & 624   & 8.81 \\
    \hline
    \textbf{GT} & 8     & 48    & 16.67 \\
    \hline
    \textbf{BUFG} & 9     & 1128  & 0.8 \\
    \hline
    \textbf{MMCM} & 1     & 22    & 4.55 \\
    \end{tabular}%
    }
  \label{tab:RESOURCE UTILIZATION PERCENTAGE}%
\end{table}%
The design utilizes 8 of the 48 high-speed serial transceivers available on the XCKU085 FPGA. These transceivers are tasked with receiving data sampled by the EV12AQ600 ADC and facilitating its transmission to the FPGA via the ESIstream protocol. In the current configuration, which includes a single ADC device, the incorporation of additional ADCs into the system would necessitate an eightfold increase in the number of GigaTransceivers (GTs) allocated for each newly added ADC. For example, integrating four EV12AQ600 ADCs into the digital beamformer system and establishing their connection to the XCKU085 FPGA through the ESIstream protocol would result in a total allocation of 32 GTs for this purpose. This expansion leaves 16 GTs available for the transmission of processed data from the digital beamformer to subsequent processing stages.

\subsection{Power Consumption and Time Analysis}
During operational testing of the full hardware setup as detailed, conducted on the EV12AQ60X-ADX-EVM evaluation board, the power consumption was recorded at 1780 mA. This figure includes the power requirements for initializing and configuring the integral components of the system: the EV12AQ600 ADC, the LMX2592 PLL, and the ESIstream protocol mechanisms. When these systems are operated in isolation, distinct power consumption values were observed: the XCKU085 FPGA, embodying the proposed design, drew 847 mA; the LMX2592 PLL, when activated and tuned to a 6.4 GHz operational frequency, required 111 mA; and the EV12AQ600 ADC accounted for 675 mA of the total power consumption. It is important to highlight the significant power demand of the ADC device, which becomes a critical consideration in the context of design expansion involving additional RF input channels and the integration of more EV12AQ600 ADC units.

From a timing perspective, the static timing analysis of the implemented design yielded the worst-case negative slack values for Setup, Hold, and Pulse Width parameters as 0.043 ns, 0.030 ns, and 0.301 ns, respectively.

\section{COMPARISON WITH STANDARD ARCHITECTURE IMPLEMENTATION}
\begin{figure}[t]
\centerline{\includegraphics[width=21pc]{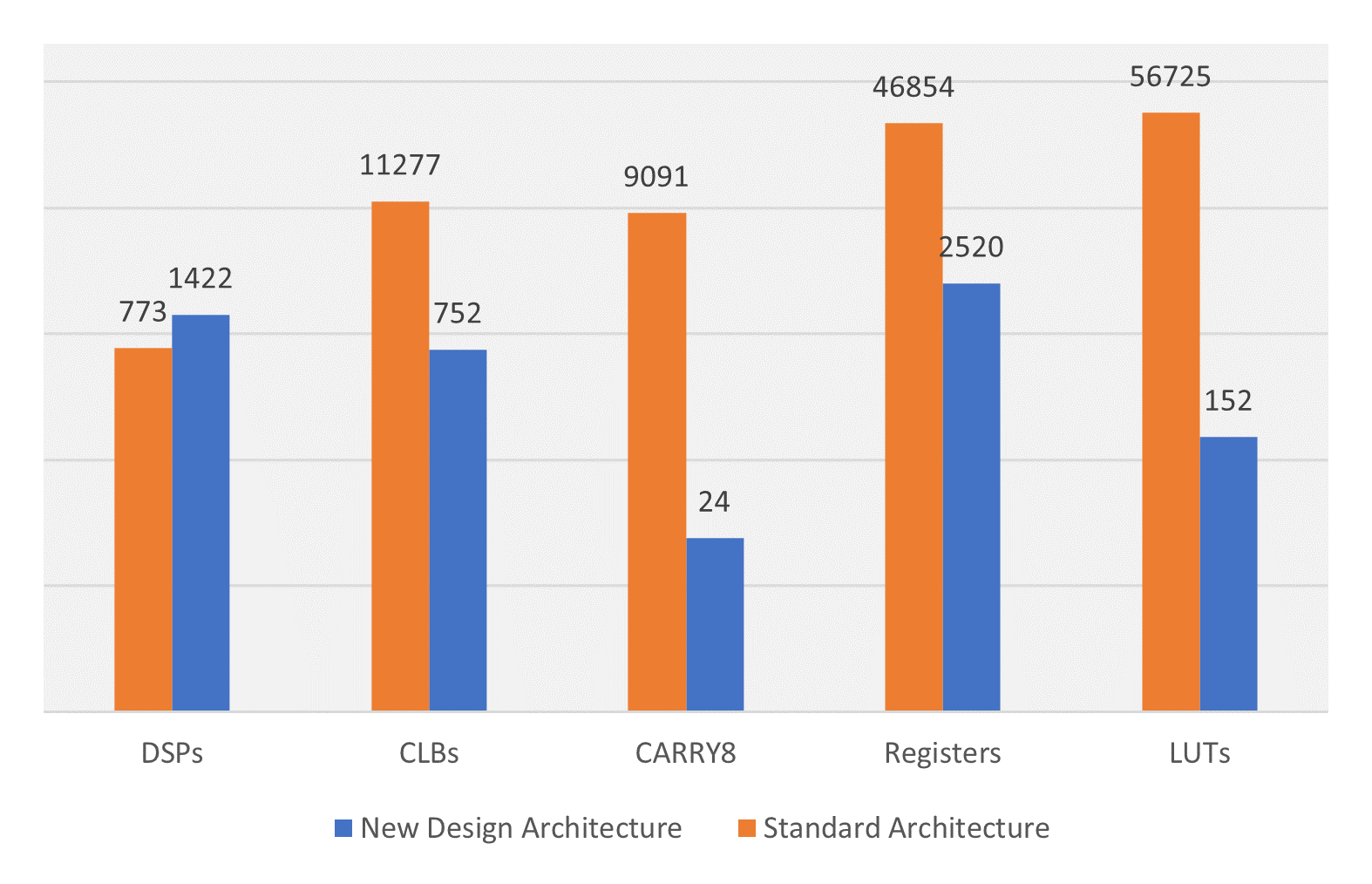}}
\caption{Resource comparison between designs.}
\label{fig:Resource comparison between designs}
\end{figure}
To enable a detailed comparison between the newly proposed architecture of the digital beamformer receiver and the conventional architecture in terms of resource utilization, power consumption, and timing analysis, both designs were implemented on the same hardware platform: the EV12AQ60X-ADX-EVM evaluation board, as depicted in Fig. \ref{fig:DBF-RX-example}.

In the conventional architecture, the signal processing flow begins with the complex down-conversion stage immediately following the EV12AQ600 ADC, succeeded by the low-pass FIR filter, and culminating in the Digital Beamformer stage. For the purpose of this comparative analysis, RF modulation processes are omitted, with the ADC performing under-sampling to concurrently digitize the signal and convert it to an intermediate frequency. This standard configuration exhibits higher resource consumption due to the operational complexity required to process each of the four input channels from the ADC separately. In contrast, in the architecture proposed within this study, the complex down-conversion and FIR low-pass filtering are applied solely to a single channel post-linear combination, emanating from the Digital Beamformer.

Additionally, the digital beamformer stage in the standard design is tasked with processing a complex input signal, thereby increasing the computational demands at this juncture. 
The resource utilization for this standard design in every stage is shown in Table \ref{tab:RESOURCE UTILIZATION PER STAGE IN STANDARD DESIGN}:
\begin{table}[htbp]
  \centering
  \caption{RESOURCE UTILIZATION PER STAGE IN STANDARD DESIGN}
    \resizebox{\columnwidth}{!}{
    \begin{tabular}{p{5.2em}|c|c|c|c|c}
    \textbf{Entity} & \textbf{LUTs} & \textbf{Registers} & \textbf{CARRY8} & \textbf{CLB} & \textbf{DSPs} \\
    \hline
    Down-conversion & \multirow{2}{*}{0}     & \multirow{2}{*}{118}   & \multirow{2}{*}{0}     & \multirow{2}{*}{55}    & \multirow{2}{*}{0} \\
    \hline
    FIR Filter & 52613 & 43980 & 8611  & 10532 & 645 \\
    \hline
    Digital \newline{}Beamforming & \multirow{2}{*}{4112}  & \multirow{2}{*}{2756}  & \multirow{2}{*}{480}   & \multirow{2}{*}{690}   & \multirow{2}{*}{128} \\
    \hline
    Total & 56725 & 46854 & 9091  & 11277 & 773 \\
    \end{tabular}%
    }
  \label{tab:RESOURCE UTILIZATION PER STAGE IN STANDARD DESIGN}%
\end{table}%
Table \ref{tab:RESOURCE UTILIZATION PER STAGE IN STANDARD DESIGN} clearly illustrates that, contrary to initial expectations, the DSP slice utilization from the XCKU085 FPGA in the standard design is markedly higher than in the newly proposed architecture. This increase in DSP slice consumption significantly detracts from the overall system performance. The primary factor contributing to this discrepancy is the extensive number and intricacy of operations required by the standard architecture, which exceed the DSP resources available within the XCKU085 FPGA. Consequently, this design necessitates a greater reliance on Logic Utilization (LUTs), registers, and Configurable Logic Block (CLB) components. Such a configuration leads to elevated power consumption and exerts a negative impact on timing across the critical path.

Table \ref{tab:RESOURCE UTILIZATION PERCENTAGE IV} provides an overview of the total number of resources used in the standard architecture design within the XCKU085 FPGA.

\begin{table}[htbp]
  \centering
  \caption{RESOURCE UTILIZATION PERCENTAGE}
    \resizebox{\columnwidth}{!}{
    \begin{tabular}{l|c|c|c}
    \textbf{Resource} & \textbf{Utilization} & \textbf{Available} & \textbf{Utilization \%} \\
    \hline
    \textbf{LUT} & 64327 & 497520 & 12.93 \\
    \hline
    \textbf{LUTRAM} & 16    & 267840 & 0.01 \\
    \hline
    \textbf{FF} & 52978 & 995040 & 5.32 \\
    \hline
    \textbf{BRAM} & 93    & 1620  & 5.74 \\
    \hline
    \textbf{DSP} & 773   & 4100  & 18.85 \\
    \hline
    \textbf{IO} & 55    & 624   & 8.81 \\
    \hline
    \textbf{GT} & 8     & 48    & 16.67 \\
    \hline
    \textbf{BUFG} & 9     & 1128  & 0.8 \\
    \hline
    \textbf{MMCM} & 1     & 22    & 4.55 \\
    \end{tabular}%
    }
  \label{tab:RESOURCE UTILIZATION PERCENTAGE IV}%
\end{table}%
Fig. \ref{fig:Resource comparison between designs} delineates the resource utilization differences, encompassing Logic Utilization (LUTs), registers, CARRY8s, Configurable Logic Blocks (CLBs), and Digital Signal Processors (DSPs), between the innovative architecture proposed in this study and the conventional design approach.

The power consumption for the hardware implementation of the proposed design, as measured on the EV12AQ60X-ADX-EVM evaluation board, registered at 1850 mA. This figure accounts for the operational states of the EV12AQ600 ADC, LMX2592 PLL, and ESIstream protocol systems, all of which were fully activated and configured.

Static timing analysis of the proposed design yielded slack values of 0.052 ns for Setup, 0.030 ns for Hold, and 0.294 ns for Pulse Width. In comparison, the standard design necessitated the integration of additional registers to satisfy timing constraints, thereby introducing greater hardware complexity and adversely influencing performance metrics.

The architecture introduced in this paper significantly optimizes resource utilization and power consumption when contrasted with the conventional design. The standard model is characterized by an increased count of LUTs and Flip-Flops (FFs), which escalates complexity while detrimentally impacting power efficiency and timing. Within the proposed design, all operations related to Digital Beamforming, including multiplications and additions, are efficiently conducted using DSP slices, thus exploiting this technology to achieve superior performance levels. Conversely, the standard design does not exclusively rely on DSP slices for these operations, leading to notable performance degradation.

Furthermore, the extension of the system to include additional RF input channels and ADCs within the standard framework mandates the incorporation of separate complex down-conversion and FIR filter stages for each new channel. This requirement results in a proportional escalation in resource demand and implementation intricacy, thereby influencing overall system performance and the capacity to adhere to timing constraints.

\section{VALIDATION}
\begin{figure}[t]
\centerline{\includegraphics[width=20pc]{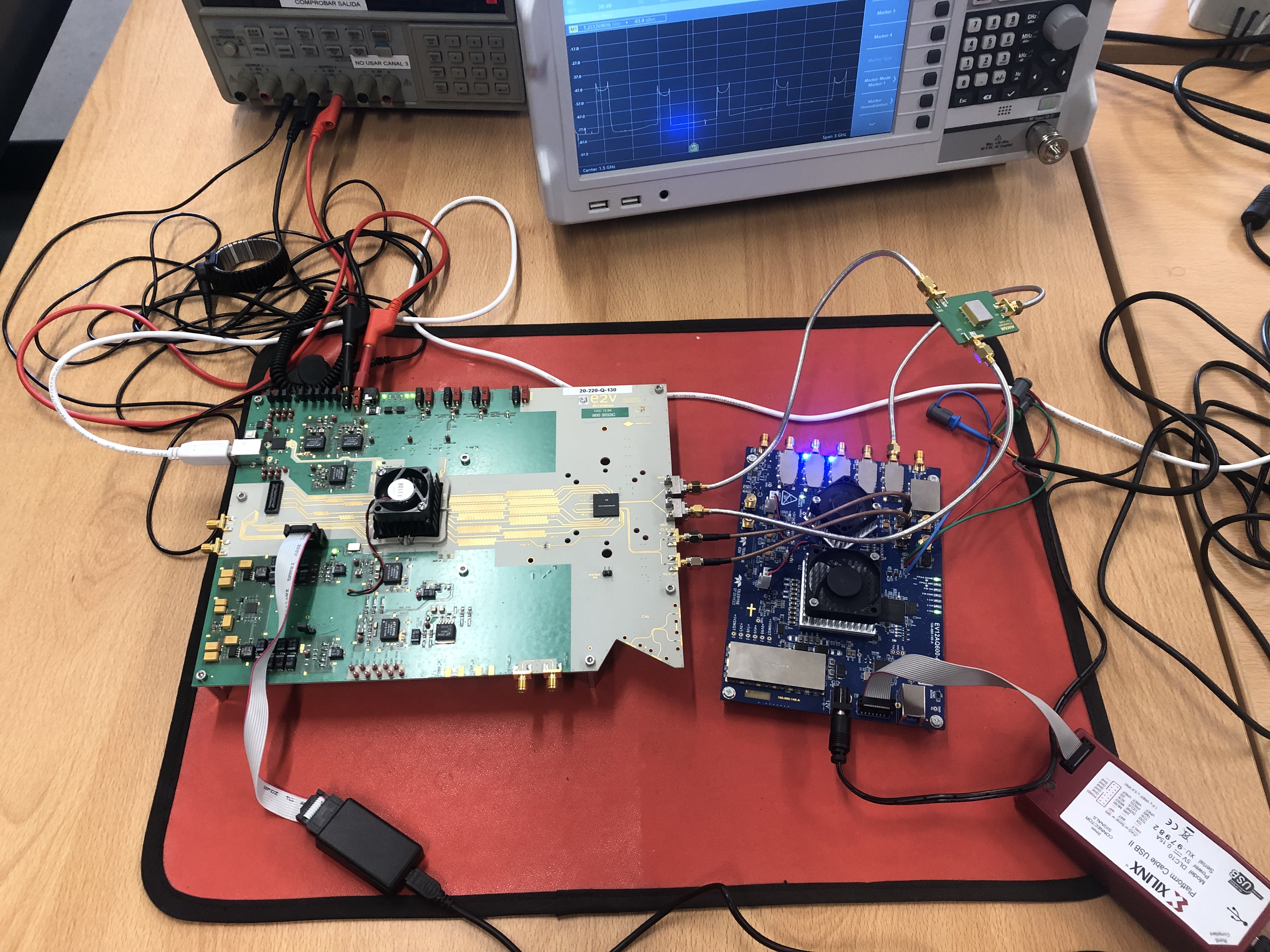}}
\caption{Test set configuration with the EV12AQ60X-ADX-EVM evaluation board.}
\label{fig:Test set configuration with the EV12AQ60X-ADX-EVM evaluation board}
\end{figure}

\begin{figure}[t]
\centerline{\includegraphics[width=20pc]{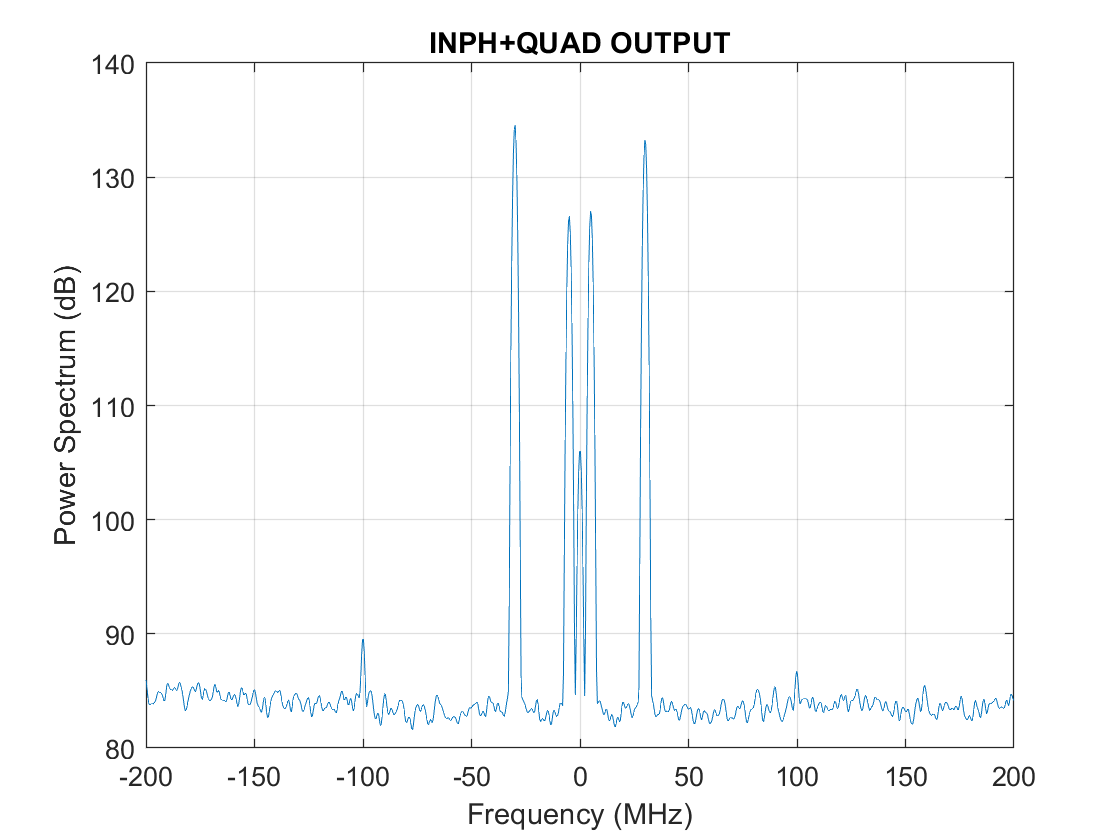}}
\caption{IQ Demodulated signal.}
\label{fig:IQ Demodulated signal}
\end{figure}

To rigorously evaluate the hardware design and functionality of the Digital Beamforming receiver implemented on the EV12AQ60X-ADX-EVM evaluation board, a series of tests involving various RF input signals were conducted. These tests aimed to verify the fidelity of the data stream output by the Digital Beamformer receiver in accurately reflecting the processed signals. For debugging and validation purposes, additional hardware modules were developed to capture and store the processed data flow, including the filtered in-phase and quadrature baseband signals, which were subsequently transmitted to a PC via a UART interface. The interaction between the EVM and the PC, necessary for triggering signal waveforms and managing data storage/retrieval, was facilitated through a suite of Python scripts. These scripts also played a crucial role in activating and configuring the EV12AQ600 ADC and the LMX2592 PLL to the required channel mode and sampling frequency. Fig. \ref{fig:Test set configuration with the EV12AQ60X-ADX-EVM evaluation board} outlines the test set configuration, illustrating the connection of the EV12AQ60X-ADX-EVM board to various instrumentation for RF signal generation and analysis. In the experimental setup, the R\&S SM300 Signal Generator is employed to deliver an input IQ modulation signal. This signal is characterized by a carrier frequency of 2 GHz, incorporating an in-phase component with a frequency of 30 MHz, and a quadrature component at 5 MHz. Fig. \ref{fig:IQ Demodulated signal} illustrates the power spectrum observed at the output of the integrated DBF and DDC processing chain.

\begin{figure}[t]
\centerline{\includegraphics[width=20pc]{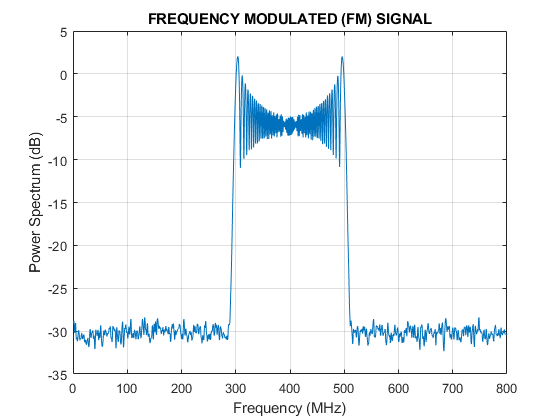}}
\caption{Frequency modulated input signal.}
\label{fig:Frequency modulated input signal}
\end{figure}

\begin{figure}[t]
\centerline{\includegraphics[width=21pc]{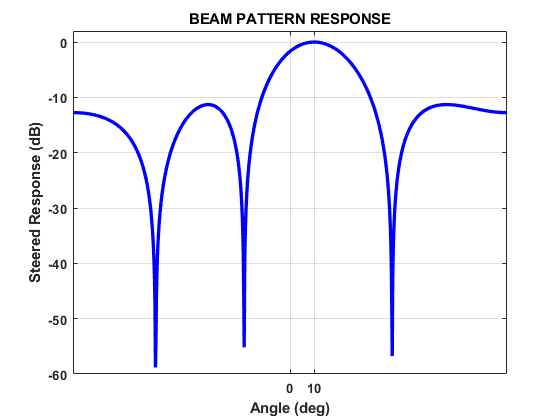}}
\caption{Beam pattern response.}
\label{fig:Beam_pattern}
\end{figure}

To emulate varying angles of arrival of the incoming RF signals to the antenna array, a Matlab script was devised. This script generates four distinct digital signal waveforms, complete with corresponding time delays for each input channel, which are then stored within the FPGA's RAM memory blocks. These simulated scenarios served to confirm the anticipated operational performance of the digital beamforming receiver, including its down-conversion processing capabilities. Fig. \ref{fig:Frequency modulated input signal} presents the power spectrum of a linear FM-modulated signal, transmitted at 3.6 GHz and down-converted to the 1st Nyquist Zone (400 MHz), featuring a base signal tone of 1 MHz and a frequency deviation of ±100 MHz, with thermal noise power set to 0 dB. The signal, arriving at a 10º angle perpendicular to the antenna array, was processed using beamforming coefficients derived from the steering vector, as depicted in the beampattern shown in Fig. \ref{fig:Beam_pattern} for an array comprising four elements. This test signal was instrumental in validating the correct functioning of the digital beamformer, complex down-conversion, and FIR filter stages within the evaluation board's new design.

\begin{figure}[t]
\centerline{\includegraphics[width=20pc]{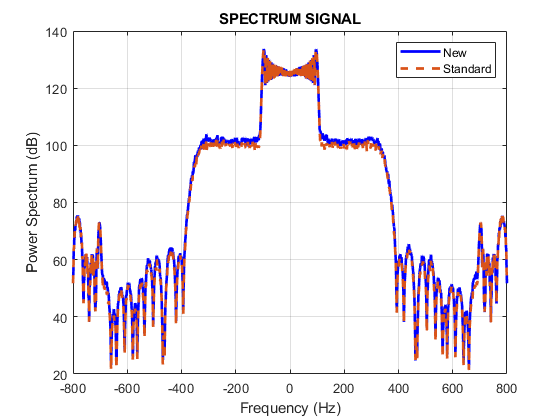}}
\caption{Output processed data.}
\label{fig:Output processed data}
\end{figure}
The resultant signal from the Digital Beamformer receiver, representing the baseband filtered beam, was captured in a RAM module within the FPGA and relayed to a PC through UART. This processed signal is visually depicted in Fig. \ref{fig:Output processed data} with a blue line, illustrating the derived characteristics of the baseband FM signal as per the newly proposed design. Concurrently, the output signal from a conventional architecture is represented by a red dashed line in the same figure. The comparison of both signals demonstrates identical characteristics, thereby underscoring the effectiveness of the novel design approach.

\section{CONCLUSION}
The development of a novel Digital Beamforming Receiver design, specifically engineered for space applications, has been realized on the EV12AQ60X-ADX-EVM Evaluation Board. This implementation leverages the synergy between the high-speed, 12-bit, quad-channel EV12AQ600 ADC and the high-performance Xilinx Kintex UltraScale XCKU085 FPGA, demonstrating an effective integration of advanced technological components. Distinctively, the design introduces the beamforming processor prior to the complex down-conversion stage, enabling a pre-emptive linear combination of input channels within the beamformer. This approach facilitates the processing of a singular data stream, comprising both real and imaginary components, through the subsequent DDC and low-pass FIR filtering stages. This strategic arrangement yields a significant reduction in both resource utilization and power consumption, establishing a precedent over traditional design methodologies.

This architecture exhibits optimized resource management, utilizing fewer LUTs and FFs while extensively employing the DSP48E2 blocks within the XCKU085 FPGA. The reduction in system complexity enhances the digital signal processing capability, effectively addressing power consumption and timing constraint considerations. Furthermore, the design is inherently scalable, allowing for the seamless addition of antenna channels. This scalability is achieved by implementing an individual Digital Beamformer stage for each additional channel, while a single DDC and FIR filter stage serve the entire system, thereby avoiding substantial escalations in resource usage or design complexity.



\bibliographystyle{elsarticle-num} 
 \bibliography{DBF_RX.bib}




\end{document}